\documentclass[11pt,english]{article}
\usepackage{amsfonts, amsmath, amssymb, verbatim, setspace, pdfsync, graphicx}
\usepackage{comment}
\usepackage{color}
\usepackage{pdfsync}
\definecolor{darkgreen}{rgb}{0,0.5,0}
\definecolor{darkblue}{rgb}{0,0,0.6}
\definecolor{purple}{rgb}{0.4,0.15,0.21}
\definecolor{black}{rgb}{.2,.2,.2}
\usepackage[colorlinks=true,citecolor=darkblue,linkcolor=black,urlcolor=darkblue]{hyperref}

\DeclareMathOperator{\p}{\partial}
\DeclareMathOperator{\h}{\theta}
\DeclareMathOperator{\Tr}{Tr}

\newcommand{\be}{\begin{equation}}
\newcommand{\ee}{\end{equation}}

\newcommand{\f}{\frac}

\begin{document}
\unitlength = 1mm
\

\begin{center}


{ \LARGE {\textsc{\begin{center}A Cardy formula for holographic hyperscaling-violating theories\end{center}}}}

\vspace{0.8cm}
Edgar Shaghoulian

\vspace{.5cm}

{\it Department of Physics} \\
{\it University of California}\\
{\it Santa Barbara, CA 93106 USA}

\vspace{1.0cm}

\end{center}

\begin{abstract}

\noindent We propose a formalism for counting the microstates of a class of three-dimensional black holes which are not asymptotically AdS. The formalism rests on the invariance of a dual field theory under a generalized modular transformation and is extended to Rindler horizons by a singular limit. We also obtain logarithmic corrections.

\end{abstract}

\pagebreak
\setcounter{page}{1}
\pagestyle{plain}

\setcounter{tocdepth}{1}

\tableofcontents

\section{Introduction}\label{intro}
Nonextremal black holes have a two-dimensional Rindler space in their near-horizon limit. The case that is better understood from a microstate point of view is that of extremal black holes, which have a near-horizon AdS$_2$. In the cases where the AdS$_2$ can be uplifted to an AdS$_3$, we have a robust microscopic derivation of the black hole entropy through the Cardy formula \cite{Cardy:1986ie, Strominger:1997eq}. An open question regards extending this type of reasoning to different near-horizon geometries, including Rindler. 

The black hole geometries we will explore in this paper can be written as \cite{Huijse:2011ef}
\be\label{blackhole}
ds^2=\ell^2r^{2\h}\left(-\f{(1-(r/r_h)^{1+z-\h})dt^2}{r^{2z}}+\f{dr^2}{r^2(1-(r/r_h)^{1+z-\h})}+\f{d\phi^2}{r^2}\right).
\ee
We will consider $z=1$ in the main text and discuss the difficulties of $z\neq 1$ in the appendix. This family of spacetimes sporadically supports holographic interpretations in string theory. We will derive a generalized Cardy formula which (modulo a few assumptions) can be understood as providing a microstate interpretation of the entropy of these black holes. Our formula can be evaluated in the singular limit $\h\rightarrow 1$, which is the case where the black hole spacetime is instead pure Rindler spacetime.  The answer obtained by doing this agrees with the entropy of a Rindler horizon. 

These black holes and their zero-temperature limits (which are singular) can be obtained as solutions of Einstein gravity minimally coupled to a self-interacting scalar:
\be
I=\int d^3 x \; \sqrt{-g} \left(R - 2 (\p \phi)^2+\f{(2-\h)(1-\h)}{\ell^2}\,e^{2\h \phi \sqrt{2/(-\h (1-\h))}} \right)\,.
\ee
The scalar takes the value $\phi = -\sqrt{-\h(1-\h)/2}\; \log r$ on shell. Although we will stick exclusively to this bulk theory, we will argue below that the same techniques can be applied to any bulk theory which has the background \eqref{blackhole} as a solution.

 These black hole backgrounds are not locally the same as their zero temperature limit, as is the case for BTZ with $\h=0$. Their entropy is given by the usual Bekenstein-Hawking formula:
\be
S_{BH}=\f{A}{4G}=\f{\pi\ell r_h^{\h-1}}{2G}, \;\; \beta = \f{4\pi r_h}{2-\h}\quad\implies \quad S_{BH}=\f{\pi\ell}{2G}\left(\f{2-\h}{4\pi}\right)^{\h-1} T^{1-\h}\,.\label{bhentropy}
\ee

Our derivation will rest on four assumptions with various levels of justification. The first will be that quantum gravity in a spacetime which is asymptotically of the form 
\be\label{empty}
ds^2=\ell^2 r^{2\h}\left(\f{-dt^2+dr^2+d\phi^2}{r^2}\right)
\ee
can be described holographically by a dual quantum field theory.\footnote{This assumption has support from string theory for certain values of $\h$. Such backgrounds (with compact manifolds along for the ride) describe non-conformal branes. For example, setting $\h=-1$ in \eqref{empty} describes the (dimensionally reduced) background for a stack of $D1$-branes \cite{Dong:2012se}. Just as in the string theory examples, we assume that there is a UV completion to the theory at high energies which we are free to ignore. In other words, we restrict the full partition function to temperatures/energies below this UV crossover scale. We assume this restricted partition function controls the hyperscaling-violating phase of the theory.}  The second assumption is that the dual field theory is invariant under a generalized modular transformation, which it inherits from the generalized scale invariance in the bulk. The third assumption is that the energy of the dual field theory state is given in terms of the energy of the gravitational field at the asymptotic boundary  through a holographically renormalized Brown-York stress-energy tensor. The final assumption will be an identification of the ground state for the field theory spectrum in terms of a bulk soliton geometry, which we will show has a negative mass and is therefore separated from the black hole spectrum by a gap. 

The steps we take in the derivation are justified for $\h<1$, with additional conceptual complications for $\h\in(0,1)$ which we will not discuss.

\subsection{Summary of results}\label{summary}
We will derive a generalized Cardy formula which reproduces the entropy of the black holes \eqref{blackhole} for $z=1$ and general $\h$, subject to positive specific heat $(1-\h) >0$. Our formula will depend on the energy of the black hole being considered, the energy of the ground state of the field theory spectrum, and $\h$.  The ground state of the field theory spectrum will be postulated in terms of a double-Wick-rotated soliton geometry in the bulk. It will have negative energy and will be separated by a gap from the rest of the spectrum. The energies of the black hole and ground state soliton will be extracted from a holographically renormalized Brown-York stress-energy tensor. 

The difficulty of the case $z\neq 1$ is discussed in the appendix, where we argue that there seems to be no field theory derivation of a generalized modular invariance or Cardy formula, even for $\h = 0$ (this is in contrast with the claims of \cite{Gonzalez:2011nz}). Nevertheless, in the appendix we will analyze the implications of such a modular invariance with $z\neq 1$ and $\h \neq 0$, were it to exist. 

We will analyze the following limits on our final answer, in nondecreasing order of pathology: (1) $\h\rightarrow -\infty$ in section \ref{littlestring}, which is a lower-dimensional analogue of the gravitational background of little string theory, (2) $\eta=-\theta/z$ fixed as $z\rightarrow \infty$ in section \ref{double}, for which the background is conformal to AdS$_2\times S^1$, (3) $z\rightarrow \infty$ in section  \ref{ads2sec}, for which the ``black hole" background becomes Rindler-AdS$_2 \times S^1$, and (4) $\h \rightarrow 1$ with $z=1$ in section \ref{flatspace}, for which the ``black hole" background becomes Rindler$\times S^1$. 

For general $z$ and $\h$, we will derive the first logarithmic correction to the black hole entropy. For $z=1$, we will employ the methods of \cite{Hartman:2014oaa} to extend the range of validity of our argument beyond the Cardy regime.

We emphasize that our generalized Cardy formula is not as powerful as the original Cardy formula for two reasons: conformal invariance allows for a universal characterization of the vacuum energy on the cylinder in terms of the central charge of the theory, and it allows one to view the Cardy formula as counting the degeneracy of local operators due to the state-operator correspondence. The lack of a state-operator map will be irrelevant for our purposes. However, if we had access to the dual field theory, an understanding of the vacuum energy in terms of an anomaly would allow a weak-coupling calculation of the vacuum energy.

\section{A generalized modular transformation}\label{genmodular}
The backgrounds \eqref{empty} admit a scaling symmetry $x_\mu\rightarrow \lambda x_\mu$, $r\rightarrow \lambda r$, and  $g_{\mu\nu}\rightarrow \lambda^{-2\h}g_{\mu\nu}$. This is just the statement that scaling is a conformal isometry of these backgrounds, and its implications for holography were first  emphasized in \cite{Kanitscheider:2008kd}. We can recast the rescaling of the metric in terms of the curvature scale, $\ell\rightarrow \lambda^{-\h} \ell$. In fact, the entire action
\be
I = \f{1}{16\pi G} \int_{\mathcal{M}} d^3x\,\sqrt{-g}\left( R-2(\p \phi)^2-V[\phi]/\ell^2\right)+\f{1}{8\pi G}\int_{\p \mathcal{M}} d^2 x\,\sqrt{-\gamma} \left(K+v[\phi]/\ell\right)
\ee 
(which includes the Gibbons-Hawking boundary term $K$ and a counterterm $v[\phi]$ we will encounter in section \ref{renorm}) is invariant under this generalized scale transformation when evaluated on-shell. This is not some coincidence of this specific action; since the Einstein tensor has an exact scaling isometry (without rescaling $\ell$), this fixes the on-shell stress-energy tensor to have the same isometry, which thereby fixes the on-shell matter Lagrangian to have the same scaling weight as the on-shell Ricci scalar (see section 2.1 of \cite{Shaghoulian:2013qia} for more details). The measure $\sqrt{g} \,d^3x$ is invariant under the transformation as well.\footnote{Although the action chosen here is special, as it follows from ``generalized dimensional reduction'' of a higher-dimensional AdS gravity \cite{Kanitscheider:2009as}, any action which can be written in the form of Einstein gravity minimally coupled to matter will have the necessary scaling isometry as argued above and in \cite{Shaghoulian:2013qia}. Thus, this form of the action is illustrative, and the results of this paper can be generalized to other theories. Conformally coupled actions with these geometries as solutions can be found in e.g. \cite{Hasanpour:2011ji}. Higher curvature terms can also be understood as simply changing the relationship between $E$ and $\ell$, e.g. for purely $n^{\textrm{th}}$ order gravity one finds $E\sim \ell^{3-2n}$, so one takes $E\rightarrow (2\pi/\beta)^{(2n-3)\h}E$ (see next page of the main text). This gives the correct result for certain quadratic and cubic theories of gravity as shown in \cite{Bravo-Gaete:2015wua}.}

The dual field theory should thus be invariant under $x_\mu \rightarrow \lambda x_\mu$ and $\ell \rightarrow \lambda^{-\h}\ell$. This rescaling of $\ell$ changes the theory, and is like rescaling $N$.\footnote{We will continue to talk about ``the" dual field theory even though the generalized scale transformation changes the theory.} If the bulk geometry is connected to AdS geometries in the UV and the IR, then the holographic c-function in the intermediate region scales as $c(r) \sim r^{\h}$, and one can understand the above symmetry in terms of a rescaling of this c-function $c(r)\rightarrow \lambda^{-\h} c(\lambda r) = c(r)$. 

Now let us consider our dual quantum field theory on a rectangular Euclidean torus (vanishing angular potential). The usual argument for invariance under $\beta \rightarrow 4\pi^2/\beta$ for CFTs comes from showing that a torus with cycles of length $\beta$ and $2\pi$ is equivalent--through a rotation, scale transformation, and translation--to a torus with cycles of length $4\pi^2/\beta$ and $2\pi$. Since these transformations are symmetries of a CFT, the theory is unchanged. In our case, these steps can be repeated with the scale transformation being replaced by a generalized scale transformation. Since one scales by $2\pi/\beta$ in this argument, we need to supplement this with $\ell\rightarrow (2\pi/\beta)^{-\h}\ell$. 

We now write the partition function of our dual field theory:
\be
Z(\beta) = \sum e^{-\beta E} g(E)\,,
\ee
where $g(E)$ is the degeneracy at energy $E$. The dependence on $\ell$ is buried in the energies $E$. Determining the exact dependence for all the energies is a hopeless task. Luckily, the only regime relevant to an application of the Cardy formula is the sector which scales linearly in $\ell$. These are gravitational states in the bulk and their linear scaling with $\ell$ is the highest possible power. For economy of notation, we will therefore rescale all the energies $E\rightarrow (2\pi/\beta)^{-\h} E$. It will become clear by the end of the derivation that the complicated $\ell$-dependence of generic states does not play a role since the partition function will project to a state which scales linearly in $\ell$. We therefore write our symmetry as
\be
\sum \exp\left(-\beta E\right)g(E) =\sum \exp\left(-\f{4\pi^2}{\beta}\left(\f{2\pi}{\beta}\right)^{-\h} E\right) g\left(\left(\f{2\pi}{\beta}\right)^{-\h } E\right)\,.
\ee
This transformation has $\beta=2\pi$ as the self-dual point. We will soon see that in the bulk this is precisely the temperature at which a Hawking-Page transition occurs. We emphasize that this symmetry relates our theory on a spatial circle at  inverse temperature $\beta$ to a \emph{different} theory on the same spatial circle at inverse temperature $4\pi^2/\beta$. 

Taking $\beta\rightarrow 0$ projects the right-hand-side to the vacuum state $E_{\textrm{vac}}$. Note that the vacuum energy is a Casimir energy, i.e. it has a dependence on the size of the compact $S^1$ which cannot be mimicked by the addition of a constant term to the action and is therefore physical. We will see later that the vacuum state has negative energy and scales linearly with $\ell$; this justifies our cavalier rescaling of all the energies. We are also assuming $g(E_{\textrm{vac}})=1$. Thus the partition function projects to
\be
Z(\beta) \approx \exp\left(-\f{4\pi^2}{\beta}\left(\f{2\pi}{\beta}\right)^{-\h} E_{\textrm{vac}}\right) \,. \label{partition}
\ee
This can be extended to twisted tori, i.e. nonvanishing angular potential, but we will not consider that case. 

 We extract the microcanonical degeneracy in a saddle-point approximation $-E/E_{\textrm{vac}}\gg 1$ as 
\be
g(E)=\int d\beta \;Z(\beta)\,e^{\beta E}\implies S =2\pi(2-\h)\, \left(\f{E\,}{1-\h}\right)^{\f{1-\h}{2-\h}}\left(-E_{\textrm{vac}}\right)^{\f{1}{2-\h}}\,.\label{cardynew}
\ee
One can check that the saddle-point temperature is indeed large. In terms of the temperature 
we have
\be
S= -E_{\textrm{vac}} (2\pi)^{2-\h}T^{1-\h}(2-\h)\,.
\ee

Before we can apply \eqref{cardynew}, we need to first identify a suitable ground state for our theory, and we need a way to calculate the energies of the black hole and the ground state. Calculating the energies requires performing holographic renormalization on our bulk theory. We should also ensure that the black hole dominates the canonical ensemble at sufficiently large temperature. The next two sections are devoted to these issues.

\section{Holographic renormalization}\label{renorm}
To calculate the mass of our black hole solutions \eqref{blackhole}, we will use the method of adding local counterterms to regulate the divergences of the Brown-York stress-energy tensor near the boundary $r_c\ll 1$ \cite{Brown:1992br, Balasubramanian:1999re}. Recall that the Brown-York tensor is given as 
\be
8\pi G\, T^{\mu\nu} = \f{2}{\sqrt{-\gamma}}\,\f{\delta I}{\delta \gamma_{\mu\nu}}=K^{\mu\nu}-K \gamma^{\mu\nu}= \ell(1-\h) r_c^{-2+\h}\eta^{\mu\nu}
\ee
for the pure hyperscaling-violating background. $K^{\mu\nu}$ and $\gamma^{\mu\nu}$ are the extrinsic curvature and induced metric of the $r=r_c$ hypersurface, and $\eta^{\mu\nu}$ is the usual Minkowski metric. The coordinates $\mu$ and $\nu$ only run over the two-dimensional hypersurface. We will interpret $T^{\mu\nu}$ as the expectation value of the stress-energy tensor in the dual hyperscaling-violating field theory, i.e. $\langle T^{\mu\nu}\rangle$ for the theory on the cylinder. 

We need to add a local counterterm $I_{ct}$ such that its functional derivative gives a contribution which cancels the divergence. Such a counterterm is given by 
\be
 I_{ct} =\f{ (\h-1)}{8\pi G \ell}\int \sqrt{-\gamma} \,e^{- \phi\sqrt{2\h/(\h-1)}}\;.
\ee
For $\h=0$ this is the usual counterterm in AdS$_3$. Recalling that the scalar field solution is given by $\phi=\sqrt{\h(\h-1)/2}\,\log r$, this gives
\be
8\pi G\, T^{\mu\nu}_{ct}=-\ell(1-\h)r_c^{-2+\h}\eta^{\mu\nu}\,.
\ee
Thus, the total stress-energy of the background vanishes. These same counterterms were derived from the point of view of ``generalized dimensional reduction'' \cite{Kanitscheider:2009as}.

\subsection{Black hole}
With the appropriate counterterm at hand, we can calculate the mass of the geometry \eqref{blackhole}:
\be
8\pi G \,  T_{tt}= \f{\ell(1-\h)}{2r_h^{2-\h}}\implies M_{bh} =\int_0^{2\pi} d\phi\, T_{tt}=\f{\ell(1-\h)}{8 G r_h^{2-\h}}\;.\label{bhmass}
\ee
This mass is consistent with the first law and gives a Smarr relation of the following form:
\be
\f{2-\h}{1-\h}\, M_{bh} = TS\,.
\ee
Notice that the Smarr relation is modified by $\h$ due to the modified scaling dimensions. 

\subsection{Ground state soliton}
We now  produce a soliton geometry which will serve as the ground state of our spectrum in the next section. Our soliton is constructed by the same double analytic continuation which produces the AdS soliton of \cite{Horowitz:1998ha}. We will henceforth assume that this soliton provides the energy of the ground state of the \emph{strongly coupled} theory.

The metric of our soliton is given by
\be
ds^2= \ell^2 r^{2\h}\left(-\f{dt^2}{r^2}+\f{dr^2}{r^2(1-(r/\tilde{r}_h)^{2-\h})}+\f{(1-(r/\tilde{r}_h)^{2-\h})d\phi^2}{r^2}\right),
\ee
where $\tilde{r}_h = (2-\h)/2$ has no interpretation as a horizon. The coordinate ranges are $0 \leq \phi < 2\pi$, $-\infty < t < \infty$ and $0<r\leq \tilde{r}_h$. The geometry as $r\rightarrow \tilde{r}_h$ looks like flat spacetime. The gauge choice made in writing down the metric makes it clear that it was produced from the black hole geometry by continuing $t\rightarrow i\phi$ and $\phi\rightarrow it$, with $r_h$ fixed by keeping a $2\pi$-periodicity in the angle $\phi$ without producing a conical deficit. This spacetime is globally static and nonsingular for $\h\leq 0$.  For $\h=0$ the curvature scalars are constant (the black hole metric in this case is just BTZ) and for $\h=1$ the curvature scalars vanish (the black hole metric is just 2D Rindler cross a unit circle). 

Using our counterterm renormalization of the previous section, the mass of this soliton is computed to be 
\be
M_{\textrm{sol}}=-\f{\ell}{8G}\left(\f{2}{2-\h}\right)^{2-\h}.\label{solmass}
\ee
Notice that this gives $M_{\textrm{sol}}=-\ell/8G$ for $\h=0$, which is the correct ground state for AdS$_3$.  For general negative $\h$, the energy $M_{\textrm{sol}} \in (-\ell/8G, 0)$, with $M_{\textrm{sol}}\rightarrow 0$ as $\h\rightarrow -\infty$. 

\section{Hawking-Page transitions}
In this section we compute the renormalized on-shell Euclidean actions of the black hole and the thermal soliton. The thermal soliton is obtained by analytically continuing the soliton to Euclidean signature and compactifying the Euclidean time coordinate with period $\beta$. This has the interpretation of a thermal gas in an asymptotically hyperscaling-violating background. We will see that at large temperature the black hole always dominates, which is necessary for our Cardyesque derivation of the entropy to make sense.

The Euclidean action we need to compute is
\be
I_E = -\f{1}{16\pi G} \int_{\mathcal{M}} d^3x\,\sqrt{g}\left( R-2(\p \phi)^2-V[\phi]/\ell^2\right)-\f{1}{8\pi G}\int_{\p \mathcal{M}} d^2 x\,\sqrt{h} \left(K-\ell^{-1}e^{-\alpha \phi}\right),
\ee
where $K$ is the trace of the extrinsic curvature of a small $r=r_c$ slice with induced metric $h_{\mu\nu}$ and $\alpha = \sqrt{2\h/(\h-1)}$. This action is a sum of on-shell terms, a Gibbons-Hawking boundary term, and a counterterm, respectively. Notice that the counterterm is the same one required to renormalize the stress-energy tensor in the previous section. We want to calculate this for the black hole background at temperature $\beta'$ and compare to our soliton with the Euclidean time coordinate compactified with period $\beta$. The physical radius of the time coordinate has to be the same between the two configurations on the $r=r_c$ slice, i.e. $\beta' (1-(r_c/r_h)^{2-\h})=\beta$. At leading order in small $r_c$ we find
\begin{align}
I_E^{\textrm{black hole}}=\f{-\beta\ell}{8 G r_h^{2-\h}}=-\f{\ell}{8G}\left(\f{4\pi}{2-\h}\right)^{2-\h}&\beta^{\h-1}\,, \quad I_E^{\textrm{thermal soliton}} = \f{-\beta\ell}{8 G}\left(\f{2}{2-\h}\right)^{2-\h}\,,\\ 
 I_E^{\textrm{black hole}}- I_E^{\textrm{thermal soliton}}&= \f{-\beta\ell}{8 G}\left(\f{1}{ r_h^{2-\h}}-\left(\f{2}{2-\h}\right)^{2-\h}\right)\,.\label{diff}
\end{align}
The mass for both configurations, computed as $M=-\p_\beta \log Z$, is consistent with the mass given by our regularized Brown-York method of the previous section. The entropy of the black hole can also be calculated as $S=\beta M+\log Z$.

For the black hole to dominate, we need \eqref{diff} to be negative, which occurs  when $r_h < (2-\h)/2$ (recall that the boundary is at small $r$, so this is a `big' black hole). In terms of the temperature, it is $\h$-independent: $\beta<2\pi$. This generalizes the case of $r_h<1$ for AdS$_3$. Also notice that as $\h\rightarrow -\infty$, the black hole can be arbitrarily small and still dominate the canonical ensemble; we will comment more on this case in section \ref{littlestring}.

In the case of AdS, it is well-known that there is no Hawking-Page transition in the Poincar\'e patch. Instead, the black brane always dominates the canonical ensemble. We can see that the same is true here by calculating the Euclidean action of an empty hyperscaling-violating background with non-compact $\phi$, which vanishes. The black string background (i.e. \eqref{blackhole} with non-compact $\phi$) has a negative Euclidean action, so it always dominates. 

\section{Applying the generalized Cardy formula}
Using the ground-state energy computed in the previous section $E_{\textrm{vac}} = M_{\textrm{sol}}$, we can write our generalized Cardy formula as 
\be
S=4\pi\left(\f{\ell}{8G}\right)^{\f{1}{2-\h}}\left(\f{E}{1-\h}\right)^{\f{1-\h}{2-\h}}.
\ee
Plugging in the mass of our black hole solution \eqref{bhmass}, we precisely reproduce the bulk Bekenstein-Hawking entropy \eqref{bhentropy}.

\subsection{Logarithmic corrections}\label{logarithmic}
Once we know that our high-temperature partition function projects onto the vacuum state, we can use this to get the degeneracy of states beyond leading order. To see this, it will be more useful to recast our derivation in a slightly different form. We begin with the partition function written as 
\be
Z(\tau,\bar{\tau})=\sum q^{E_L+E_{\textrm{vac}}/2} \bar{q}^{E_R+E_{\textrm{vac}}/2} g(E_L, E_R), \qquad q=e^{2\pi i \tau}\,.
\ee
We have shifted the energies so that $E_L+E_R+E_{\textrm{vac}}$ is the energy on the cylinder. We will again specify to $\tau=\tau_1+i\tau_2 = i\tau_2$, i.e. set the angular potential to zero. The degeneracy is extracted from the partition function as
\be
g(E)=\int d\tau\; Z(\tau) e^{-2\pi i \tau(E+E_{\textrm{vac}})}\,.
\ee
We define $\tilde{Z}(\tau)$ through $Z(\tau)=\tilde{Z}(\tau)e^{2\pi i \tau E_{\textrm{vac}}}$ and use the generalized modular invariance to write  
\be
g(E)=\int d\tau_2 \;\tilde{Z}\left(\tau_2^{-1+\h }\right) e^{2\pi \tau_2 (E+E_{\textrm{vac}})} e^{-2\pi E_{\textrm{vac}}\tau_2^{-1+\h}}\,,\label{integrand}
\ee
where we have assumed that the vacuum state scales linearly with $\ell$. In the limit $-E/E_{\textrm{vac}}\rightarrow \infty$ we can evaluate this integral by a saddle point approximation, keeping the first subleading correction.  The saddle-point value of $\tau_2$ is given as 
\be
\tau_2^s=\left(\f{-E_{\textrm{vac}}(1-\h)}{E+E_{\textrm{vac}}}\right)^{\f{1}{2-\h}}\,.
\ee
It is important that $\tilde{Z}$ is a sum of exponentially suppressed contributions at the saddle $\tau_2^s$, except for the contribution of the ground state $E=0$ (recall that we shifted the energies so that $E+E_{\textrm{vac}}$ is the energy of a state on the cylinder). Thus we have $\tilde{Z}\left[(\tau_2^s)^{-1+\h }\right]=1$. 

The leading contribution is given by evaluating the integrand of \eqref{integrand} at $\tau_2^s$. The first correction is given by expanding around $\tau_2^s$ to quadratic order and performing the resulting Gaussian integral:
\be
g(E) =\f{ \left(-E_{\textrm{vac}}(1-\h)\right)^{\f{1}{4-2\h}} E^{\f{\h-3}{4-2\h}}}{\sqrt{2-\h}}\;\exp\left(\f{2\pi E(2-\h)}{1-\h} \left(\f{-E_{\textrm{vac}}(1-\h)}{E}\right)^{\f{1}{2-\h}}\right).
\ee
The entropy can be written as 
\be
S=\f{A}{4G} - \f{3-\h}{2(1-\h)} \log \f{A}{4G}+\dots\,.
\ee
The prefactor of the logarithm is a universal number which generalizes the $-3/2$ of BTZ black holes \cite{Carlip:2000nv}. It is negative for positive specific heat $\h<1$, consistent with general arguments in \cite{Das:2001ic}. This procedure can be extended to arbitrary order \cite{Loran:2010bd}. 

\subsection{Extension of range of validity}\label{hartman}
We can now ask about extending the range of validity of our Cardy formula. Phenomenologically, our formula gives the correct answer for a black hole of any size in the bulk, even though the formula was derived for asymptotically large black holes. We should expect that the validity of the formula breaks down below the Hawking-Page transition, since the black hole no longer dominates the canonical ensemble in that regime. In the formalism of \cite{Hartman:2014oaa}, we can split up our bulk states into light, medium, and heavy states:
\be
\textrm{light}: E_{\textrm{vac}}\leq E \leq \epsilon, \qquad \textrm{medium}: \epsilon < E < (1-\h)(-E_{\textrm{vac}}), \qquad \textrm{heavy} : E\geq (1-\h)(-E_{\textrm{vac}})\,.
\ee
The lower bound on the heavy range is defined as the energy at which the black hole dominates the canonical ensemble. The methods of \cite{Hartman:2014oaa} can then be applied to this parameterization, and we list a few of the results here. We omit details since the technical steps are equivalent.

The Cardy formula is universal in the heavy range as long as the spectrum of light states is bounded as 
\be
\rho(E)\lesssim \exp[2\pi(E-E_{\textrm{vac}})]\,.
\ee
Modulo a genericity assumption explained in \cite{Hartman:2014oaa}, the medium-energy range satisfies the bounds
\be
4\pi\left(\f{\ell}{8G}\right)^{\f{1}{2-\h}}\left(\f{E}{1-\h}\right)^{\f{1-\h}{2-\h}}\lesssim S(E) \lesssim 2\pi( E-E_{\textrm{vac}})\,.
\ee
We can further uses these results to make sharp statements about the spectrum of the theory, such as the existence of certain states, but we will not pursue this here.

\section{Two intriguing limits}
We consider our formalism in the limit of vanishing and infinite effective dimensionality $d_{\textrm{eff}}=1-\h$. The former limit $\h=1$ corresponds to flat space, whereas the latter $\h\rightarrow -\infty$ is a lower-dimensional version of the background dual to little string theory \cite{Shaghoulian:2013qia}. We will find that at vanishing effective dimensionality the entropy is governed entirely by the ground state, whereas at infinite effective dimensionality the entropy is governed entirely by the state of the black hole being considered. This can be seen immediately from our generalized Cardy formula $S \sim E^{\f{1-\h}{2-\h}}\,(-E_{\textrm{vac}})^{\f{1}{2-\h}}$. 
\subsection{Flat space}\label{flatspace}
The geometry being considered becomes flat in the limit $\h\rightarrow 1$. Notice that the effective dimensionality heuristic $d_{\textrm{eff}}=1-\h$  in this case indicates that the dual theory should be $(0+1)$-dimensional, i.e. a matrix quantum mechanics.

The geometry in this situation is a patch of flat spacetime and the ``black hole" geometry is simply diffeomorphic to two-dimensional Rindler space times a circle:
\be
ds^2=-(1-r/r_h)dt^2+\f{dr^2}{1-r/r_h}+d\phi^2=-\f{u^2}{4r_h^2}dt^2+du^2+d\phi^2, \qquad u^2=4r_h^2(1-r/r_h)\,.
\ee
Our entropy and ground state formulas all have well-defined limits in this case, and we find
\be
E_{\textrm{vac}}= -\f{1}{4G}, \qquad S=2\pi (-E_{\textrm{vac}}) =  \f{\pi}{2G}\,.
\ee
Notice that this reproduces the usual entropy formula for Rindler in terms of the dimensionless Rindler energy \cite{Susskind:1993ws}. The soliton geometry in this case is simply ordinary flat space in polar coordinates: 
\be
ds^2=-dt^2+\f{dr^2}{1-2r}+(1-2r)d\phi^2=-dt^2+du^2+u^2d\phi^2,\qquad u^2=1-2r\,.
\ee
The only reason this is assigned a mass is because the stress-tensor is evaluated at $r=0$ ($u=1$); while $r=0$ is the conformal boundary in the $\h<0$ geometries, for $\h=1$ we can extend the coordinate to $r\rightarrow -\infty$, i.e. $u\rightarrow \infty$. Notice also that the enigmatic range vanishes in this limit.

\subsection{Very little string theory}\label{littlestring}
Our entropy formula \eqref{cardynew} is linear in the energy as $\h\rightarrow -\infty$:
\be
S=2\pi E\,. \label{littleentropy}
\ee
This indicates a Hagedorn density, with inverse Hagedorn temperature $\beta = 2\pi$. Since we want the limit of our formula instead of the limit of the final answer, we did not plug in the expressions for $E$ and $E_{\textrm{vac}}$ before taking the limit. It is not clear that such a limit makes sense, but we will now obtain this result without a limiting procedure.

 The metric can be written as 
\be
ds^2=\f{\ell^2}{r^2}\left(-dt^2+\f{dr^2}{r^2}+d\phi^2\right)\,,
\ee
which shows that the scaling transformation is $r\rightarrow\lambda r$, $\ell\rightarrow \lambda \ell$. The inherited symmetry of the boundary theory is a rescaling of $\ell$ by an arbitrary amount, with the boundary coordinates left untouched. This is a remarkable symmetry if true. To implement it, we need to operate slightly differently than in section \ref{genmodular}. Since the theory will not project to its ground state, we write:
\be
\int dE \,e^{-\beta E} \rho(E)= \int  dE \,\lambda \,e^{-\beta \lambda E}\rho\left(\lambda E\right)\implies \rho(E) =e^{\beta E}(A/E+B\delta(E))\,.
\ee 
The symmetry automatically implies a Hagedorn density for the sector of gravitational states $E\sim \ell$, including a subleading correction. We did not use such a continuous description of the partition function in cases where we project to the ground state since the unique, gapped ground state would lead to the vanishing of the density near $E_{\textrm{vac}}$. The subleading corrections take the same form as the density of states following from the symmetries of AdS$_2$ \cite{Jensen:2011su, almheiri}. 

Let us compare to the bulk black hole directly. For reasons that will become clear, we will choose an angular periodicity $\phi\sim \phi+4\pi$:
\be
ds^2=\f{\ell^2}{r^2}\left(-(1-r/r_h)dt^2+\f{dr^2}{r^2(1-r/r_h)}+d\phi^2\right)\,.
\ee
The thermodynamic quantities are given as
\be
 \beta = 4\pi, \quad S=\f{\pi\ell}{Gr_h}, \quad M=\f{\ell}{4Gr_h}, \quad \implies\quad M = TS\,,
\ee
where the mass is calculated by the usual Brown-York method at $r_c\ll 1$. Notice that the temperature is independent of the horizon as required for a Hagedorn spectrum. 

It is clear that our formula  $S=2\pi E$ is off by a factor of two. However, this is an illusion due to the fact that our formalism was derived for a torus with angular periodicity $2\pi$. Since the bulk black hole can only exist at temperature $4\pi$, this will imply that the analogous ground state soliton exists at angular periodicity $4\pi$. This means that to handle this case we would have to repeat our field theory analysis on a torus with angular periodicity $4\pi$ along the spatial cycle; this would give us precisely the factor of two needed to agree with the bulk.

\section{Outlook}\label{conclusions}
Spacetimes which are conformal to anti-de Sitter space with a pure power-law conformal factor are ubiquitous in string theory as the backgrounds sourced by non-conformal branes. We have considered a family of such backgrounds and their finite-temperature generalizations in three dimensions. The generalized scale invariance of the bulk geometries was shown to imply a generalized modular invariance of the dual field theory on a torus, which was used to derive a formula for the asymptotic degeneracy of states. This formula was checked explicitly against the entropy of the bulk black hole solutions and precisely agrees. The formula also reproduces the entropy of a Rindler horizon in the limit $\h\rightarrow 1$. The case of dynamical critical exponent $z\neq 1$ is discussed in the appendix.

There are many directions left open by this work:
\begin{itemize}
\item Construct distinct black hole solutions to distinct actions with the same asymptotics and test the formula \eqref{cardynew} in those cases.
\item Generalize this formalism to include angular momentum and test the generalized entropy formula for rotating black holes.
\item Check the prediction of logarithmic corrections against other methods.
\item Use the self-dual point of our generalized modular transformation $\beta=2\pi$ to further constrain the theory as in \cite{Hellerman:2009bu}.
\item Prove a ``positive" energy theorem for the spacetimes \eqref{empty} which shows that the energy is bounded below by the soliton. Since we are in three bulk dimensions, the boundary topology is a cylinder and we need not worry about complications inherent to Wick-rotated  solitons in higher dimensions \cite{Horowitz:1998ha}.
\item  Provide a field theory argument for the ground state energy on the cylinder. Without this or a positive energy theorem, our derivation is incomplete.
\item Perform a generalized asymptotic symmetry group analysis to see if the generalized scale invariance enhances to a more constraining, possibly infinite-dimensional symmetry in the boundary theory.
\item Derive a universal formula for entanglement entropy from the field theory.  Such derivations often use local symmetries at key steps in the argument, whereas we have only used a global scale invariance. 
\item Provide a technical justification of our calculation of the Rindler entropy. 
\end{itemize}

\section*{Acknowledgments}
I would like to acknowledge useful discussions with Ahmed Almheiri, Sean Hartnoll, Gary Horowitz, Raghu Mahajan, Joe Polchinski, Alexandre Streicher, and Tomonori Ugajin. This work is supported by NSF Grant PHY13-16748.

\section{Appendix}\label{appendix}
Lifshitz field theories have an anisotropic scale invariance which has been used to argue for a generalized modular invariance. This invariance  projects the partition function at high temperatures to the vacuum state \cite{Gonzalez:2011nz}. At the moment, we do not believe that these arguments are correct. The ``isomorphism" of Lifshitz algebras used in \cite{Gonzalez:2011nz} does not allow one to swap thermal and spatial cycles as is assumed. This swap is a key step in deriving modular properties and Cardy-like formulas. In the simplest cases it follows from Euclidean rotational invariance (i.e. Lorentzian boost invariance, which is a symmetry of the theories considered in this paper). This is absent in Lifshitz field theories. In fact, it is clear that even if one could swap cycles, it would lead to a relation between a theory with dynamical critical exponent $z$ and a theory with dynamical critical exponent $1/z$. This cannot be interpreted as an invariance of a single theory. The best-case-scenario is to relate two different theories and argue that the high-temperature partition function of the theory with critical exponent $z$ is determined solely by the vacuum energy of the related theory with critical exponent $1/z$, but even this argument does not have support. The only possibility we see is to rewrite the partition function as a trace over a different Hilbert space
\be
Z(\beta) = \Tr e^{-\beta H} = \Tr e^{-2\pi J(\beta)}\approx e^{-2\pi J_{\textrm{min}}}\,.
\ee
In the final step we have assumed the spectrum of the angular momentum operator is bounded below and projected to this minimum angular momentum by taking $\beta\rightarrow 0$, i.e. $2\pi/\beta\rightarrow \infty$. Recall that $J$ is quantized in units of $\beta$ in the transformed partition function, and notice that $J_{\textrm{min}}$ is \emph{not} the angular momentum of the vacuum state. The general Lifshitz theory will not have bounded angular momentum, nor would such a formula agree with the subsequent black hole ``checks" of the Cardy formula of \cite{Gonzalez:2011nz}. This brings us to the next point: we do not believe reproducing the entropies of bulk black holes, performed in \cite{Gonzalez:2011nz}, is a convincing check of the formula. This is due to the fact that the equivalence (up to rescaling) of the Euclidean actions of the ground-state soliton and the black hole ensures, through the success of Euclidean gravity methods for computing black hole entropy, that the  energy of the soliton contains the same data as the entropy of the black hole.\footnote{Of course, one could similarly complain that the entropy calculations in the present paper are not convincing checks of our formula, but in our case Lorentz invariance allows rigorous derivations of the generalized Cardy formula in both this paper and--with different assumptions--in \cite{Shaghoulian:2015kta}.}

Nevertheless, we will analyze the implications of such a generalized modular invariance for Lifshitz theories, in case there is a legitimate derivation of such a symmetry. For the general hyperscaling-violating Lifshitz field theory, we can therefore project the partition function at high temperatures to its vacuum state by the formalism in section \ref{genmodular}:
\be
\sum \exp(-\beta E)\,g(E)\approx \exp\left(-(2\pi)^{1+1/z}\beta^{-1/z}\left(\f{2\pi}{\beta}\right)^{-\h/z} E_{\textrm{vac}}\right).\label{cardylif}
\ee
This follows from the fact that in the pure Lifshitz field theory, the scaling transformation required for the projection is $\phi\rightarrow (2\pi/\beta)^{1/z}, \,\tau\rightarrow (2\pi/\beta)\tau$. Notice that we need $(1-\h)/z >0$ to relate high temperatures to low temperatures. This is the requirement in the bulk of positive specific heat. The microcanonical degeneracy becomes
\be
S=\f{2\pi E\,(1-\h+z)}{1-\h} \left(\f{-E_{\textrm{vac}}(1-\h)}{zE}\right)^{\f{z}{1-\h+z}},
\ee
which can be written in terms of the temperature as
\be
S=-E_{\textrm{vac}}(2\pi)^{\f{1-\h+z}{z}}\left(\f{1-\h+z}{z}\right)T^{(1-\h)/z}\,.
\ee
These expressions agree with the ones in the main text for $z=1$ and the ones in \cite{Gonzalez:2011nz} for $\h=0$. As advertised in \ref{summary}, we can trade $\h$ and $z$ for $n\equiv (1-\h)/z$ to find 
\be
S=2\pi E\,\f{1+n}{n} \left(\f{-E_{\textrm{vac}}\,n}{E}\right)^{\f{1}{1+n}}  =(2\pi)^{1+n}(1+n)T^n\,.
\ee
The energies $E$ and $E_{\textrm{vac}}$ will however depend explicitly on both $z$ and $\h$ and cannot be traded for $n$. 

To test these formulas, we will need energies for the ground state and the black hole. Since we do not have a field theory argument for the energy of the ground state, we will again assume that the ground state configuration in the bulk is given by a double Wick rotation of the black hole geometry. 

Rather than perform holographic renormalization on these spacetimes to extract the masses, we will instead cheat. We will extract the mass of the black hole geometry by using $dM=TdS$. We will use this mass to deduce the Euclidean on-shell action for the configuration through $M=\p_\beta I^{on-shell}$ (up to an irrelevant constant). Recall now that the Euclidean thermal soliton is given by a Wick rotation of this black hole, with $r_h$ fixed by $\beta(\tilde{r}_h)=2\pi$ to maintain a $2\pi$ periodicity in the angle $\phi$. This means that the thermal soliton has the same Euclidean on-shell action as the black hole with $r_h=\tilde{r}_h$ (see e.g. \eqref{diff}; one factor of $\beta$ needs to be extracted to account for the temperature of the thermal soliton). Thus, we will differentiate this on-shell action for the thermal soliton to extract its mass, which equals the mass of the soliton with non-compact time. Performing these manipulations, we find
\be
M_{sol}=-\f{\ell z}{8G}\left(\f{2}{1-\h+z}\right)^{\f{1-\h+z}{z}},\qquad M_{bh} = \f{\ell(1-\h)}{8Gr_h^{1-\h+z}}\,.
\ee
Using the solitonic mass as the ground state energy of the field theory and plugging into \eqref{cardylif}, we precisely reproduce the Bekenstein-Hawking formula for the bulk black hole entropy. It would be useful to extract the masses of the black hole and soliton in a more honest way, such as in \cite{Ayon-Beato:2015jga}, as a test of our formalism.

\subsection{AdS$_2$ limit}\label{ads2sec}
In the limit $z\rightarrow \infty$ the geometry becomes asymptotically AdS$_2 \times S^1$. We see this by taking the limit after the following redefinitions:
\be
r^{2z}=\tilde{r}^2,\quad  r_h^{2z}=\tilde{r}_h^{2z}, \quad \ell=\tilde{\ell}\,z, \quad t=\tilde{t}/z, \quad  \phi = \tilde{\phi}/ z\,.
\ee
Notice that this is an infinite redefinition of $t$ and $\phi$. This means that any periodicity in $\tilde{\phi}$ or Euclidean time diverges. Dropping the tildes, the metric becomes 
\be
ds^2=-\ell^2\left(\f{-(1-r/r_h)}{r^2}dt^2+\f{dr^2}{r^2(1-r/r_h)}+d\phi^2\right)\,.
\ee
Performing the additional redefinitions $r\rightarrow 2 r_h$, $t\rightarrow 2r_h$, $(1-r/r_h)/r^2 = y^2$, and absorbing $\ell$ into the coordinates, we see that this is just Rindler-AdS$_2$ cross a ``circle": 
\be
ds^2=-(y/\ell)^2 dt^2+\f{dy^2}{1+(y/\ell)^2}+d\phi^2\,.
\ee
We formally keep the periodicity of the circle as $2\pi \ell z$ even though $z=\infty$. Notice that for any $\ell$ our entropy formula in the limit $z\rightarrow \infty$ gives 
\be
S=-2\pi E_{\textrm{vac}}\,, \qquad E_{\textrm{vac}}=-\f{\ell z}{4G}\,.\label{ads2}
\ee
An additional factor of $z$ was inserted (honestly) to account for the new (infinite) periodicity of $\phi$. This is the same as the factor of $2$ which was necessary in section \ref{littlestring}. Notice that the limit $\ell\rightarrow \infty$ gives flat Rindler space, analyzed in section \ref{flatspace}, whereas the limit $\ell\rightarrow 0$ connects to AdS$_2 \times S^1$ in the Poincar\'e patch. This latter limit is the case most relevant for near-horizon limits of extremal black holes. 


\subsection{Double-scaling limit}\label{double}

In the double-scaling limit $\h\rightarrow -\infty$ and $z\rightarrow \infty$ with $\eta=-\h/z$ fixed \cite{Hartnoll:2012wm}, the microscopic entropy becomes
\be
S=2\pi E\; \f{1+\eta}{\eta}\left(\f{-E_{\textrm{vac}}\, \eta}{E}\right)^{\f{1}{1+\eta}} = -E_{\textrm{vac}}(1+\eta)(2\pi)^{1+\eta}\,T^{\eta}\,.\label{cardydouble}
\ee
Notice that  $\eta\rightarrow \infty$ recovers $S=2\pi E$ as in \eqref{littleentropy}, whereas $\eta \rightarrow 0$ recovers $S=-2\pi E_{\textrm{vac}}$ as in \eqref{ads2}. This agreement is necessary, since those limits correspond to $\h\rightarrow -\infty$ and $z\rightarrow \infty$, respectively. 

Performing redefinitions similar to the previous section, the black hole background in this limit becomes
\be
ds^2=\f{\ell^2}{r^{2\eta}}\left(-\f{\left(1-(r/r_h)^{1+\eta}\right)dt^2}{r^2}+\f{dr^2}{r^2\left(1-(r/r_h)^{1+\eta}\right)}+d\phi^2\right)
\ee
with thermodynamic data
\be
T= \f{1+\eta}{4\pi r_h}, \qquad M=\f{\eta\ell}{8G r_h^{1+\eta}}, \qquad S = \f{\pi\ell}{2G r_h^\eta}\,,
\ee
where the mass was determined from $dM=TdS$ as in the previous section. We can also extract the soliton mass using the same trick as in the previous section:
\be
M_{\textrm{sol}} = -\f{\ell}{2^{2-\eta}(1+\eta)^{1+\eta} G}\;.
\ee
Plugging into the formula \eqref{cardydouble}, we again find agreement with the bulk entropy. In this case, we have simply dropped the factor of $z^2$ which appears in the $g_{rr}$ term when taking the limit (which we scaled into the coordinates in the previous section); this is to illustrate that our results are independent of this factor. 

Interestingly, the Cardy formula, all of the bulk thermodynamic data, the emblackening factor, and the soliton mass for this case equal the case in the main text with $z=1$ upon performing the identification $\eta \equiv 1-\h$, even though the geometries are distinct. The case of $\eta=1$ maps to the CFT case $\h=0$. We can therefore write our generalized Cardy formula for $\eta=1$ as the usual Cardy formula for CFTs, $S=2\pi \sqrt{c\,E/3}$ with $c=3\ell/2G$. This is related to the fact that these geometries uplift to AdS$_3$ \cite{Gubser:2009qt}.

\subsection{Logarithmic corrections}
We can also extract the leading logarithmic correction in this case. By the same procedure explained in section \ref{logarithmic}, we find 
\be
\rho(E)=\f{(-E_{\textrm{vac}}(1-\h))^{\f{z}{2(1-\h+z)}} (E z)^{-\f{1-\h+2z}{2(1-\h+z)}} |z|}{\sqrt{1-\h+z}} \exp\left(\f{2\pi E\,(1-\h+z)}{1-\h} \left(\f{-E_{\textrm{vac}}(1-\h)}{zE}\right)^{\f{z}{1-\h+z}}\right)\,.
\ee
We can use this to write the entropy as 
\be
S=\f{A}{4G}-\f{1-\h+2z}{2(1-\h)} \log \f{A}{4G}\,.
\ee
The prefactor is again universal. We can trade it for the scaling of the entropy $S\sim T^n$ with $n=(1-\h)/z$ to get $-(2+n)/(2n)$. This is negative for positive specific heat, consistent with general arguments in \cite{Das:2001ic}. 

\small{
\bibliography{hyperscalingcardybiblio}
\bibliographystyle{apsrev4-1long}}

\end{document}